\documentclass[useAMS,usenatbib]{mn2e}

\usepackage{graphicx}                   
\usepackage{float}
\usepackage{amsmath}
\usepackage{amssymb}
\usepackage{multirow}
\usepackage{bm}
\usepackage{color}

\input{epsf.sty}                        
\input{psfig.sty}
\begin{document}

\title[Pulsar Glitches in a Strangeon Star Model]
{Pulsar Glitches in a Strangeon Star Model}

\author[Lai et al.]{X. Y. Lai$^{1,2}$, C. A. Yun$^{1,3}$, J. G. Lu$^{4,5}$, G. L. L$\ddot{u}$$^3$, Z. J. Wang$^3$, R. X. Xu$^{4,5}$\\
$^1$Xinjiang Astronomical Observatory, Chinese Academy of Sciences, Urumqi 830011, China\\
$^2$School of Physics and Engineering, Hubei University of Education, Wuhan 430205, China\\
$^3$School of Physics, Xinjiang University, Urumqi 830046, China\\
$^4$School of Physics, Peking University, Beijing 100871, China\\
$^5$Kavli Institute for Astronomy and Astrophysics, Peking University, Beijing 100871, China}

\maketitle

\begin{abstract}
Pulsar-like compact stars provide us a unique laboratory to explore properties of dense matter at supra-nuclear densities.
One of the models for pulsar-like stars is that they are totally composed of ``strangeons'', and in this paper we studied the pulsar glitches in a strangeon star model.
Strangeon stars would be solidified during cooling, and the solid stars would be natural to have glitches as the result of starquakes.
Based on the starquake model established before, we proposed that when the starquake occurs, the inner motion of the star which changes the moment of inertia and has impact on the glitch sizes, is divided into plastic flow and elastic motion.
The plastic flow which is induced in the fractured part of the outer layer, would move tangentially to redistribute the matter of the star and would be hard to recover.
The elastic motion, on the other hand, changes its shape and would recover significantly.
Under this scenario, we could understand the behaviors of glitches without significant energy releasing, including the Crab and the Vela pulsars, in an uniform model.
We derive the recovery coefficient as a function of glitch size, as well as the time interval between two successive glitches as the function of the released stress.
Our results show consistency with observational data under reasonable ranges of parameters.
The implications on the oblateness of the Crab and the Vela pulsars are discussed.
\end{abstract}

\begin{keywords}
dense matter - stars: neutron - pulsars: general
\end{keywords}

\section{Introduction}
\label{Introduction}

The state of matter in pulsar-like compact stars depends on non-perturbative quantum chromodynamics (QCD) which is challenging in fundamental particle physics.
Tremendous efforts have been tried to solve this problem, though no consensus has been achieved.
Pulsar-like compact stars could be strange quark stars instead of neutron stars, if strange quark matter (composed of nearly equal numbers of deconfined $u$, $d$ and $s$ quarks) in bulk constitutes the true ground state of strong-interaction matter, as stated by the so-called Witten's conjecture~\citep{Witten1984}.
However, at realistic densities of pulsars, i.e. $\rho\sim 2-10 \rho_0$ ($\rho_0$ is the saturated nuclear matter density), the effect of non-perturbative QCD would be very significant, and the state of matter is far from certainty.

From astrophysical points of view, the matter composed of strange quark-clusters could form when baryonic matter is compressed by the huge gravity in the process of supernova explosion.
Although no calculation or simulation has been performed to verify such kind of state, this could be understood phenomenologically from both top-down and bottom-up scenarios.
In the top-down scenario, starting from the deconfined quark matter with the inclusions of stronger and stronger interaction between quarks, one could get strangon matter where the quarks are grouped into ``quark-clusters''.
On the other hand, in the bottom-up scenario, starting from the hadronic state, the strangeness may play an important role in gigantic nuclei so that the degree of freedom would not be nucleons but quark-clusters with strangeness.

With nearly equal numbers of $u$, $d$ and $s$ quarks, a strange quark-cluster is also called a ``strangeon'' as an abbreviation of ``strange nucleon'' (with strangeness degree of freedom).
Stars composed of strangeons are called ``strangeon stars'' (in some previous papers, it is called strange quark-cluster star).
Based on phenomenological analysis and comparison with observations, the strangeon star model is proposed~\citep{Xu2003,LX2009}.
Similar to traditional quark stars, strangeon stars have a large amount of strange quarks with the number nearly equals to that of up and down quarks.
However, strangeon stars are composed of strangeons, distinguished from traditional quark stars which are composed of deconfined quarks.

Different manifestations of pulsar-like compact objects have been discussed previously (see a review by~\cite{LX2017} and references therein) in the strangeon star model.
Strangeon stars could be bare, but may have negligible atmospheres~\citep{Wang2017}.
Because strangeons are non-relativistic, and the residual colour interaction between them may have a short-distance repulsion core (an analogy of that between nucleons), the equation of state of strangeon matter could be very stiff so that the observed massive pulsars~\citep{Demorest2010,Antoniadis2013} could be naturally expected~\citep{LX2009,LGX2013, GLX2014}.
When the temperature is significantly low ($<$ 1 MeV), strangeon stars could solidify~\citep{Dai2011}, and the gravitational energy released in starquakes of solid strangeon stars could power the radiation of anomalous X-ray pulsars (AXPs) and soft gamma-ray repeaters (SGRs)~\citep{Xu2006,Tong2016}.
Moreover, the recently observed gravitational waves GW170817~\citep{ligo2017} as well as the electromagnetic radiation (e.g.,~\cite{Kasliwal2017}) could be understood if the signals come from the merge of two strangeon stars in a binary~\cite{Lai2017b}.
in this paper we will focus on explaining the glitch behaviors of strangeon stars.

It is well known that glitches reflect the interior structure of pulsars, and in turn the nature of pulsars is certainly the starting point for understanding the physics of glitches.
Pulsar glitches, i.e. the sudden spin-ups of pulsars, are one type of pulsar timing irregularity and have been detected for many known pulsars.
The mechanism of glitches still remains to be well understood, although a large data set has been accumulated~\citep{Espinoza2011}.
The glitch size, often defined as the relative increases of spin frequencies during glitches $\Delta\Omega_g/\Omega$, has a bimodal distribution ranging from $\sim 10^{-10}$ to $\sim 10^{-5}$, with peaks at $\sim 10^{-9}$ and $\sim 10^{-6}$~\citep{Lyne2000,Wang2000,Yuan2010,Espinoza2011,Yu2013}.

The physics of glitches have been made in framework of neutron stars, under two main models.
The first regards glitches as starquakes of an oblate crust~\citep{Ruderman1969,Baym1971}, and the second regards glitches as the result of rapid transfer of angular momentum from inner part to the crust of a neutron star~\citep{Anderson1975}.
Although the first model has difficulty in explaining the glitch activities of the Velar pulsar, it could be the trigger for the second model~\citep{AA2017}.
The review about glitch models of neutron stars is given by~\cite{Haskell2015}.
Some properties about the glitch behaviors of neutron stars, such as the geometry of crustquake and the time time interval between two successive glitches, have been given by~\cite{Akbal2015,Akbal2017}.

In this paper, we study the glitch behaviors in a strangeon star model, based on the starquake scenario initialized by~\cite{Ruderman1969} and then developed by~\cite{Baym1971}.
In the case of neutron stars, when the star spins down, strain energy develops in its crust until the stress reaches a critical value, then a starquake occurs and some of the stress is relieved.
During the starquake of the star, the moment of inertia, $I$, of the crust suddenly decreases, so its rotation frequency suddenly increases due to conservation of angular momentum.
The strain could also be produced by magnetic and superfluid force~\citep{Link2000}, and the breaking strain of neutron stars has been discussed which is related to mountain-buiding and gravitational wave emission~\citep{Horowitz2009}.
Although the magnetic force as well as the decaying magnetic field would also produce the strain of a strangeon star, here we do not consider such effects and only consider the strain developed by spinning-down.

Pulsar glitches could be the result of starquakes of solid strangeon stars~\citep{Zhou2004,Zhou2014}, and the detailed modeling about the glitch behaviors compared with observations is the main focus of this paper.
In the case of neutron stars, when the star spins down, strain energy develops in its crust until the stress reaches a critical value, then a starquake occurs and some of the stress is relieved.
During the starquake of the star, the moment of inertia, $I$, of the crust suddenly decreases, so its rotation frequency suddenly increases due to conservation of angular momentum.
The difference between neutron stars and strangeon stars is that, the whole body of the strangeon star, rather than only the crust in Baym-Pines model, is in a solid state.
Therefore, unlike the neutron star with a superfluid core enveloped by a thin solid crust suffering the strain only, the whole body of a strangeon star has rigidity and feels the strain.

A detailed analysis of the fracture process~\citep{Baym1971} showed that, in the case of a incompressible solid star, the starquake would begin with a cracking of some places at the equatorial plane below the surface of the star and then propagate outwards.
Considering this result, we propose that, just when the starquake of a strangeon star occurs, the propagation of the cracking outwards from the equatorial plane would induce a plastic flow in the outer part of the star, and the sphere inside the outer layer which was under tension would change its shape elastically.
In other words, just when the starquake occurs, the inner motion of the star is divided into plastic flow and elastic motion, both of which would decrease the value of $I$ of the star.

The plastic flow moves tangentially in the outer layer of the star and brings some material from the equator to the poles, so our starquake scenario corresponds to the bulk-invariant case proposed by~\cite{Zhou2014}, which would not be accompanied by significant release of gravitational energy.
The bulk-variable and bulk-invariable starquakes have been proposed in the strangeon star model by~\cite{Zhou2014}.
In bulk-variable starquakes, the global radius of the star $R$ changes significantly with $-\Delta R/R \sim \Delta \Omega_g/\Omega$.
Consequently a huge amount of energy would released, which would originate glitches accompanied by X-ray bursts (e.g. that of AXPs/SGRs) even if the glitch sizes are as small as $10^{-9}$.
In bulk-invariable starquakes, on the other hand, only the oblateness of the star changes with $-\Delta \epsilon\sim \Delta \Omega_g/\Omega$, which would originate glitches without evident energy release (e.g. that of the Crab and the Vela pulsars).

It should be noted that, in the case of the bulk-invariable starquake in~\cite{Zhou2014}, the whole star is treated as an elastic body both before and during starquakes.
Differently, in this paper, although the star is still treated as an elastic body before starquakes, we introduce an plastic flow in the fractured part of the star during starquakes which does not change the volume of the star, so the starquakes we consider in this paper is of bulk-invariable.
Dividing the inner motion of the star during starquakes into the plastic flow and elastic motion, the glitch behaviors of two typical glitching pulsars, the Crab and the Vela pulsars, could be well understood.

The plastic flow in the fractured outer layer of the star redistributes the matter inside the star and would be hard to recover, and on the other hand, the elastic motion of the sphere inside the outer layer tends to recover the shape, like a stretching spring being released.
During the spin-up process, the elastic motion would also induce fracture of the sphere, which would then make the recovery process much slower than the spin-up process.
This could help us to understand the recovery behavior after glitches.
Glitch recovery could usually be fitted by an exponential decay of spin frequency with amplitude $\Omega_d$.
The recovery coefficient, often denoted by
$Q=\Delta\Omega_d/\Omega_g$, indicates the degree of recovery of
spin frequency after glitch.
Under our starquake scenario, we derive the relation between $Q$ and $\Delta \Omega_g/\Omega$ which is consistent with observational data under reasonable parameters.
We also derive the relation between the time interval of two successive glitches and the released stress, and our results are consistent with the data of two typical glitching pulsars, the Crab and the Vela pulsars.
The results implies that the actual oblateness of the Crab pulsar might be larger than that of the Velar pulsar.

The real process of starquakes should be complicated, and a simplified model of starquake proposed in this paper is certainly a crude description of the real process.
We expect that more data including glitch sizes and recovery coefficients will help us to give better constraints for the properties of strangeon stars.
%
%
%
%
%

%

\section{Starquakes of strangeon stars}

%
%
The total energy of a rotating incompressible star with mass $M$, radius $R$ and angular momentum $L$ is~\citep{Baym1971}
\begin{equation}
E=E_0+\frac{L^2}{2I}+A\epsilon^2+B(\epsilon-\epsilon^{\prime})^2, \label{energy}
\end{equation}
where $E_0$ is the total energy in the non-rotating case, $I$ is the moment of inertia, $\epsilon$ is the oblateness that relates to $I$ via $I=I_0(1+\epsilon)$ where $I_0$ is the moment of inertia of a non-rotating incompressible star,
and $\epsilon^{\prime}$ ($>\epsilon$) is the reference oblateness (the oblateness of the unstrained sphere).
The coefficient measuring the departure of gravitational energy relative to the non-rotating case is $A=-\frac{1}{5}E_{\rm grav,0}$, where $E_{\rm grav,0}=-\frac{3}{5}GM^2/R$ is the gravitational energy of the star with the uniform density in the non-rotating case, and the coefficient measuring the strain energy is $B=\mu V/2$ where $\mu$ is the mean shear modulus and $V=4\pi R^3/3$ is the whole volume of the star~\citep{Baym1971}.
The strain energy is $E_{\rm strain}=B(\epsilon-\epsilon^{\prime})^2$, where $\epsilon_0$ is the initial oblateness.
The mean stress $\sigma$ is defined as
\begin{equation}
\sigma=|\frac{1}{V}\frac{\partial E_{\rm strain}}{\partial \epsilon}|=\mu (\epsilon^{\prime}-\epsilon). ~\label{sigma}
\end{equation}
%
%

\begin{figure}
\begin{center}

 \includegraphics[width=3 in]{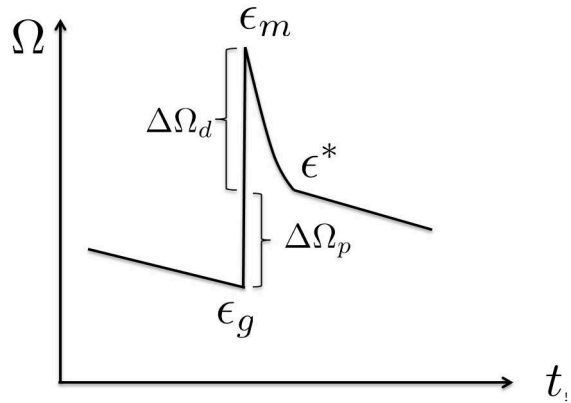}

\end{center}
\caption{
 An illustration of the oblateness $\epsilon$. The ellipticity of the star is exaggerated in this figure. The oblateness just before glitch (when the stress reaches the critical value) and just after the recovery of the glitch are denoted by $\epsilon_g$ and $\epsilon^*$ respectively.  $\epsilon_m$ is the minimal oblateness.
\label{figQt}}
\end{figure}

The process of one glitch is illustrated in Fig~\ref{figQt}.
A normal spin-down phase begins at the end of last glitch, and the elastic energy is accumulating and the stress $\sigma$ is increasing.
When the value of $\sigma$ has increased to the critical value $\sigma_c$, the star fragments and releases the elastic energy.
%
%

\subsection{The starquake process}

For a neutron star, during the cracking, the moment of inertia $I$ of the crust suddenly decreases, and its rotation frequency suddenly increase as the result of angular momentum conservation.
A solid strangeon star is naturally expected to have glitches as the result of starquakes, under the scenario similar to the neutron stars with crusts, but the difference is that, the strangeon star is totally in solid state, and when a quake occurs the whole star would be affected.

Before a detailed theoretical demonstration of our model, we should present the picture of starquake process, including the how the star cracks and how the star reacts to the cracking.
Taking a completely solid star with uniform density as an example, \cite{Baym1971} discussed the starquake process and derived some conclusions listed in the following:
(1) the starquake would begin with a cracking of the equatorial plane below the surface of the star and then propagate outwards along the equatorial plane;
(2) before the cracking the equatorial plane is under tension in the inter part ($r<(8/9)^{1/2}R$, $R$ is the radius of the star) and under compression in the outer part ($r>(8/9)^{1/2}R$);
(3) the critical strain increases towards the center of the star.
Moreover, based on the cracking picture of neutron star crust originated from \cite{Ruderman1969}, \cite{Link2000} proposed that when the starquake relaxes the stress, the shearing motion along the fault decreases the oblateness of the star and pushes matter to higher latitudes.
Although the starquake process in \cite{Baym1971} is actually not applicable to neutron stars, we could apply it to solid strangeon stars which are completely solid.

Based on the above conclusions, we propose a simple scenario of starquake process of solid strangeon stars.
Before the starquake, the whole star is an elastic body which is accumulating the elastic energy.
The critical stress first reaches at the equator plane with distance $r$ (where the matter are under tension) from the centre of the star.
After that, the starquake begins with a cracking of the equatorial plane and then propagate outwards, whereas the sphere inside radius $r$ temporarily emains unfractured.
Being under tension before starquake, after cracking the sphere inside the outer layer would behave like a released tensional spring.
In the meanwhile, the outer layer of the star breaks along fault lines, forming platelets and moving towards the poles like a plastic flow.
The plastic flow moves tangentially and brings some material from the equator to the poles.
After both of the plastic flow and elastic motion cease, i.e., after recovery of the glitch, the whole star recovers to an elastic body.

\begin{figure}
\begin{center}

 \includegraphics[width=3 in]{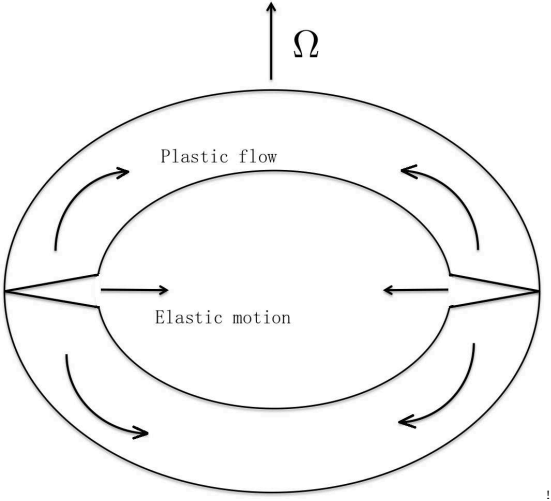}

\end{center}
\caption{
 An illustration of the starquake process in this paper. The starquake begins with a cracking of the equatorial plane and then propagate outwards~\citep{Baym1971}, whereas the sphere inside remains temporarily unfractured.
During the starquake, the sphere would undergo the elastic motion and behave like a released tensional spring, and the plastic flow induced in the fractured outer layer of the star moves tangentially and brings some material from the equator to the poles.
\label{fig_2}}
\end{figure}

Therefore, we divide the inner motion of the star during starquake into plastic flow and elastic motion respectively (shown in Fig~\ref{fig_2}), both of which would change the moment of inertia $I$ of the star.
The plastic flow in the outer layer of the star leads to the redistribution of matter, breaking the density uniformity.
The elastic motion of the sphere inside the outer layer, on the other hand, only changes the shape or oblateness of the star.
Based on such scenario, we parameterize $I$ as
\begin{equation}
I=I_0(1+\epsilon)(1+\eta), \label{I}
\end{equation}
where $\epsilon$ ($\ll 1$) is the oblateness, and $\eta$ ($\ll 1$) denotes the deviation from uniform density,
The change of $\epsilon$ reflects the elastic motion, and the change of $\eta$ reflecting the plastic flow.
%

The strain energy released in the cracking of the equatorial plane would be converted into thermal energy and kinetic energy of the plastic flow.
\cite{Ruderman1969} assumed that the entire stress is relieved in the quake, while \cite{Baym1971} suggested that only a part of the stress is released and the plastic flow is negligible.
As stated above, in this paper we consider that both the elastic and plastic motion play roles in the spin-up process.
We further assume that, when the plastic flow ceases, the kinetic energy is converted into strain energy again.
This means that the change of $\eta$, i.e. the redistribution of matter, does not lead to releasing of stress.
During starquakes, the stress is released as oblateness $\epsilon$ reduces.

It is worth noting that, most glitches are followed by an increase in the spin-down rate $|\dot{\Omega}|$ with positive values of $\Delta \dot{\Omega}/\dot{\Omega}$~\citep{Espinoza2011}.
The spin-down change accompanying the change in rotation rate at the time of glitches is expected in our starquake model.
The starquake event affect the spin-down change by reducing the moment of inertia $I$ and subsequently increasing the spin angular frequency $\Omega$.
%
%
The spin-down rate deriving from the magnetic dipole radiation depends on both $\Omega$ and $I$ in terms of $\dot{\Omega} (<0)\propto -\Omega^3/I$, so just after the spin-up epoch $|\dot{\Omega}|$ is larger than the pre-glitch value, i.e., $\Delta \dot{\Omega}<0$, giving positive values of $\Delta \dot{\Omega}/\dot{\Omega}$.

\subsection{The recovery coefficient}

A glitch is an sudden increase of angular velocity denoted by $\Delta \Omega_g$, and the recovery is usually described as the sum of $\Delta \Omega_d$ and $\Delta \Omega_p$, which are respectively the decay and persistent increase of angular velocity, $\Delta \Omega_g=\Delta \Omega_d+\Delta \Omega_p$.
The recovery coefficient $Q$ is defined as $Q=\Delta \Omega_d / \Delta \Omega_g$.
The angular momentum is conserved during a glitch, i. e. $\Delta L=\Delta(I\Omega)=0$.
From the definition of $I$ in (\ref{I}), we can get
\begin{equation}
\frac{\Delta\Omega_g}{\Omega}=-\Delta\epsilon_m-\Delta\eta, ~\label{DeltaOmegag}
\end{equation}
where the change in oblateness $\Delta \epsilon_m=\epsilon_m-\epsilon_g$ ($<0$) and $\epsilon_m$ is the minimum values of oblateness, and
\begin{equation}
\frac{\Delta\Omega_p}{\Omega}=-\Delta\epsilon-\Delta\eta, ~\label{DeltaOmegap}
\end{equation}
where the change in oblateness $\Delta \epsilon=\epsilon^*-\epsilon_g$ ($<0$), and the values of oblateness just before the glitch and just after the recovery of the glitch are denoted by $\epsilon_g$ and $\epsilon^*$ respectively.
We have taken into account the fact that, the change of $I$ due to the redistribution of matter would not recover, so the above two equations have the same term $\Delta\eta$.
Therefore, from the Eqs.(\ref{DeltaOmegap}), (\ref{DeltaOmegag}) and the definition of $Q$, we can get
\begin{equation}
Q=\frac{\epsilon^*-\epsilon_m}{\Delta\Omega_g/\Omega}.
\end{equation}

As we have stated in \S 2.1, the motion of the interior matter of the star is divided into plastic and elastic motion, so the quantity $\epsilon^*-\epsilon_m$ reflects the elastic recovery, since the change of $I$ by plastic flow would hard to recover.
The value of recovery coefficient shows whether the change of $I$ is dominated by $\Delta \epsilon$ or $\Delta \eta$.
In the former case,  the value of $\epsilon^*-\epsilon_m$ tends to be as large as $\Delta\Omega_g/\Omega$, and in the latter case, the value of $\epsilon^*-\epsilon_m$ tends to be zero.
Therefore, we can write $\epsilon^*-\epsilon_m$ as the function of $\Delta\Omega_g/\Omega$ with a parameter $a$ as (regardless of the case of $Q>1$)
\begin{equation}
\epsilon^*-\epsilon_m=\frac{\Delta \Omega_g}{\Omega}\exp\left(-a\frac{\Delta \Omega_g}{\Omega}\right).
\end{equation}
The value of $\epsilon^*-\epsilon_m$ reaches its maximal value $\sim 1/a$ when $\Delta\Omega_g/\Omega=1/a$.
The relation between $Q$ and $\Delta\Omega_g/\Omega$ is then
\begin{equation}
Q=\exp\left(-a\frac{\Delta \Omega_g}{\Omega}\right). \label{Q}
\end{equation}

Although $Q$ is simply parameterized as in Eq.(\ref{Q}), the parameter $a$ may not be a constant.
Different pulsars could have different values of $a$, depending on the mass, radius, strength of magnetic field or even the glitch sizes.
In our starquake model at the present stage, it is difficult to take into all the possible factors into account, and here we simply parameterize $Q$ as in Eq.(\ref{Q}) and estimate the range of $a$ according to observational data.

\begin{figure}
\begin{center}

  \includegraphics[width=3.5 in]{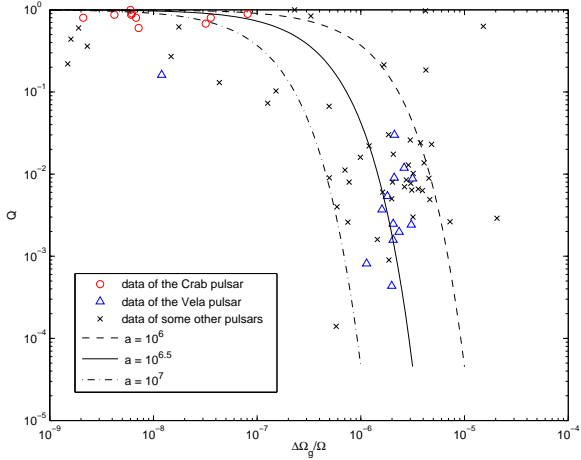}
\end{center}
\caption{
The relation between $Q$ and $\Delta\Omega_g/\Omega$ derived from Eq.(\ref{Q}) under three values of $a$ with $10^{7}$, $10^{6.5}$ and $10^{6}$ are shown in Fig~\ref{fig_Q}.
The data of the Crab and the Vela pulsars are shown by red circles and blue triangles respectively, and the data of some other pulsars who have the $Q$ data are shown by black crosses for comparison~\citep{Espinoza2011}.
\label{fig_Q}}
\end{figure}

The relation between $Q$ and $\Delta\Omega_g/\Omega$ derived from Eq.(\ref{Q}) under three values of $a$ with $10^{7}$, $10^{6.5}$ and $10^{6}$ are shown in Fig~\ref{fig_Q}.
The data of the Crab and the Vela pulsars are shown by red circles and blue triangles respectively, and the data of some other pulsars who have the $Q$ data are shown by black crosses for comparison~\citep{Espinoza2011}.
These curves show that the values of $a$ are in the range from $\sim 10^6$ to $\sim 10^7$, corresponding to the result that $\epsilon^*-\epsilon_m$ increases with $\Delta \Omega_g/\Omega$ for small values of $\Delta \Omega_g/\Omega$ and reach its maximum value in the range from $\sim 10^{-7}$ to $\sim 10^{-6}$, after which $\epsilon^*-\epsilon_m$ decreases rapidly with $\Delta \Omega_g/\Omega$.
%

Although the actual value of $\epsilon^*-\epsilon_m$ could be difficult for us to estimate, that $\epsilon^*-\epsilon_m$ has the maximum value in the range from $\sim 10^{-7}$ to $\sim 10^{-6}$ could be reasonable, if we take the consideration that it should be smaller than the oblateness of an ideal fluid star with the same rotating frequency $\Omega$, which has the configuration of Maclaurin ellipsoid with oblateness $\epsilon_{\rm Ma}=\Omega^2 I_0/4A\sim 10^{-4}$ for a typical young pulsar with $M=1.4 M_{\odot}$, $R=10$ km and $\Omega\sim 100\ {\rm rad\ s^{-1}}$, where $I_0$ and $A$ have been defined before.
On the other hand, because the  maximum value of $\epsilon^*-\epsilon_m$ should be less than the actual oblateness of the star, we could put the lower limit for the latter to be $\sim 10^{-7}$.

It is worth mentioning that, the sphere inside the fractured outer layer of the star would also be fractured partially during the elastic motion, which would make the value of the mean shear modulus much smaller than that given in Eq.(\ref{constant})~\citep{Zhou2004}.
Consequently, the recovery timescale could be much larger than the spin-up timescale~\citep{Zhou2004}.

\subsection{The time interval between two successive glitches}

Because the star is treated as an elastic body both before a glitch and after recovery of the glitch, the time interval between two successive glitches $t_q$ (from this glitch to the next) could be derived as that in ~\cite{Baym1971},
\begin{equation}
t_q\simeq \frac{|\Delta \sigma|}{\dot{\sigma}}=\frac{2A(A+B)}{BI_0}\frac{|\Delta \epsilon|}{\Omega \dot{\Omega}}, \label{tq}
\end{equation}
where $|\Delta \sigma|$ is the stress released during glitch and $\dot{\sigma}$ is the increase of stress as time, both of which can be derived from Eqs. (\ref{energy}) and (\ref{sigma}) to achieve the above equation, and $|\Delta \epsilon|= \epsilon_g-\epsilon^*$ is the persistent decrease of oblateness after glitch.
We apply the above expression of $t_q$ for strangeon stars, as we have assumed that only the reduction of oblateness, not the redistribution of matter, would lead to releasing of stress during glitches.
The constant term of can be evaluated as
\begin{eqnarray}
\frac{2A(A+B)}{BI_0}&\simeq & 5.7\times 10^{10}{\rm erg / g/cm^2} \left(\frac{\rho}{3\rho_0}\right)^{7/3}\nonumber \\&\times&\left(\frac{M}{1.4M_\odot}\right)^{2/3}
\left(\frac{10^{32} \rm erg/cm^3}{\mu}\right), \label{constant}
\end{eqnarray}
where the star is assumed to have uniform density $\rho$ ($\rho_0$ is the saturated nuclear matter density), and the value of the mean shear modulus $\mu$ have been given as in~\cite{Xu2003}.
We can see that, although $t_q$ is dependent on the mass the the star $M$, its value would not change much when $M$ increases from $0.5 M_\odot$ to $2 M_\odot$.
Parameterizing $I$ as (\ref{I}), the persistent increase of angular velocity $\Delta \Omega_p$ after glitch is related to both $\Delta \epsilon$ and $\Delta \eta$,
\begin{equation}
\frac{\Delta \Omega_p}{\Omega}=|\Delta \epsilon|+|\Delta \eta|,
\end{equation}
which means that, $|\Delta \epsilon|$ is not directly related to $\Delta \Omega_p/\Omega$, so we cannot predict the time interval between this and the next glitches $t_q$ only from the value $\Delta \Omega_p/\Omega$ of this glitch.

\begin{figure}
\begin{center}

  \includegraphics[width=3.5 in]{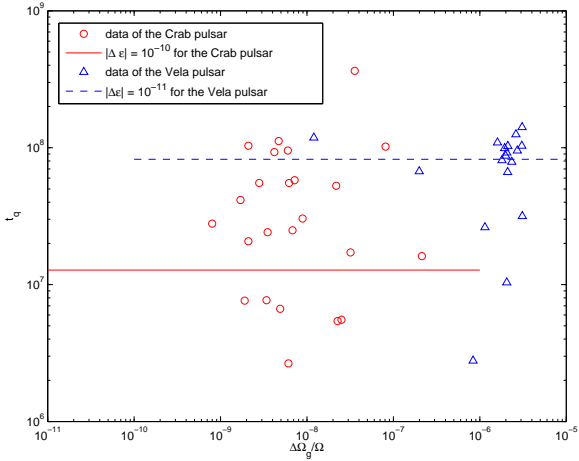}
\end{center}
\caption{
The values of $t_q$ calculated from Eq. (\ref{tq}), with $|\Delta \epsilon|=10^{-10}$ for the Crab (red line) and $|\Delta \epsilon|=10^{-11}$ for the Vela (blue dashed line) pulsars. The data for $\Delta\Omega_g/\Omega$ and $t_q$ of the Crab (red circles) and the Vela (blue triangles) pulsars are also shown~\citep{Espinoza2011}.
\label{fig_tq}}
\end{figure}

%
%
The persistent decrease of oblateness $|\Delta \epsilon|$ should generally be related to the actual oblateness of pulsars $\epsilon$, so we assume that it has the same value for each pulsar, then Eq. (\ref{tq}) shows that one particular pulsar has one explicit value of $t_q$ independent of glitch sizes.
This conclusion could be qualitatively consistent with the fact that, the Crab pulsar and the Vela pulsar have nearly the same values of $t_q$, although their glitch sizes differ by at most three orders of magnitude.
The data for $\Delta\Omega_g/\Omega$ and $t_q$ of the Crab (red circles) and the Vela (blue triangles) pulsars are shown~\citep{Espinoza2011} in Fig.\ref{fig_tq}, where the values of $t_q$ calculated from Eq. (\ref{tq}) have also been shown, with $|\Delta \epsilon|=10^{-10}$ for the Crab (red line) and $|\Delta \epsilon|=10^{-11}$ for the Vela (blue dashed line) pulsars.
%

Although setting the value $|\Delta \epsilon|$ to be independent of glitch size might be a rough approximation, we can see from the data that the dependence of $\Delta \epsilon$ on glitch size should not be significant.
In addition, the value of $|\Delta \epsilon|$ of the Crab pulsar is larger than that of the Vela pulsar by nearly one order of magnitude, which imply that, the actual oblateness $\epsilon$ of the Crab pulsar could be much larger than that of the Vela pulsar.

\subsection{Energy released during a glitch}

The energy available during a glitch includes the releasing of the gravitational energy and the strain energy.
Even for the Velar pulsar, as demonstrated below, the glitch in our starquake scenario would not be energetic enough to produce an X-ray enhancement.

During a starquake, the plastic flow is induced in the outer layer of the star and moves tangentially.
%
%
From Eq. (\ref{energy}) and \cite{Baym1971}, the gravitational energy is $E_{\rm grav}=E_{\rm grav, 0}+A\epsilon^2$, so the gravitational energy released is
\begin{equation}
\Delta E_{\rm grav}=2A\epsilon |\Delta \epsilon|,
\end{equation}
where $A=\frac{3}{25}GM^2/R$ as defined at the beginning of \S 2, $\Delta \epsilon$ is the change of oblateness during the glitch.
By assuming that the plastic flow would not lead to significant release of strain energy, and taking that the strain energy (elastic energy) is $E_{\rm strain}=B(\epsilon-\epsilon^\prime)^2$, the strain energy released is
\begin{eqnarray}
\Delta E_{\rm strain}&=&2B(\epsilon^\prime-\epsilon)(|\Delta\epsilon^\prime|-|\Delta\epsilon|)\nonumber \\ &=&2A(\epsilon^\prime-\epsilon)|\Delta\epsilon|,
\end{eqnarray}
where $\Delta\epsilon^\prime$ is the change of reference oblateness, and the second line comes from the result that $|\Delta\epsilon|=\frac{B}{A+B}|\Delta \epsilon^\prime|$ (i.e. only a part of the strain is relieved~\citep{Baym1971}).

Therefore, for a typical pulsar with $M=1.4 M_\odot$ and $R=10$ km, the energy released during a glitch is
\begin{eqnarray}
\Delta E&=&\Delta E_{\rm grav}+\Delta E_{\rm strain}=2A\epsilon^\prime|\Delta\epsilon|\nonumber\\
&\simeq& 10^{37} {\rm erg}\ \frac{\epsilon^\prime}{10^{-4}}\ \frac{|\Delta\epsilon|}{10^{-11}},
\end{eqnarray}
where the value of the reference oblateness $\epsilon^\prime$ is chosen to be the value if the star were an ideal fluid star (whose configuration is the Maclaurin ellipsoid).
For the Vela pulsar, $\Omega\sim 100\ {\rm rad\ s^{-1}}$, so $\epsilon^\prime \sim \epsilon_{\rm Ma}=\Omega^2 I_0/4A\sim 10^{-4}$ (strictly speaking, the value of $\epsilon^\prime$ should be slightly larger than that of $\epsilon_{\rm Ma}$).
We can see that, even for the Vela pulsar who has large glitch sizes, since the value of $|\Delta\epsilon|\sim 10^{-11}$ (derived in \S 2.3 and shown in Fig.~\ref{fig_tq}), the energy released during a glitch is too low to produce an X-ray enhancement.

\section{Conclusions and discussions}

The nature of pulsar-like compact stars has attracted a lot of attention since the discovery of pulsar fifty years ago but still remains to be solved.
Our previous work showed that they could actually be strangeon stars composed of strangeons (an abbreviation of ``strange nucleons'') which form due to the strong coupling between quarks.
Strangeon matter would condensate to form a solid state when the temperature is much lower than the interaction energy between strangeons.
Various observational properties of solid strangeon stars have been discussed previously, and in this paper we focus on explaining the glitch behaviors of solid strangeon stars.

Based on the starquake model established by~\cite{Baym1971} and the bulk-invariable starquakes in~\cite{Zhou2014} which would lead to glitches without significant energy releasing, we propose the starquake process of strangeon stars.
As the spinning down of the star, the strain energy develops until the stress reaches a critical value, when a starquake occurs and the some of the stress is relieved.
The starquake of the solid strangeon star, which begins with cracking of the equatorial plane and propagates outwards, would induce a plastic flow in the outer layer of the star, and the sphere inside the layer would change its shape elastically.
In this starquake process, therefore, we could divide the inner motion of star during starquake into the plastic flow and the elastic motion, both of which would change the moment of inertia of the star and have impact on the glitch behaviors.

During a starquake, the plastic flow in the fractured part of the star moves tangentially in the outer layer of the star and brings some material of the star from the equator to the poles.
The elastic motion of the sphere inside the layer, on the other hand, changes its shape and behaves like a spring being released from tension.
Therefore, the moment of inertia $I$ would first decrease to the minimum, corresponding to a sudden increase of rotation frequency, and then recover.
The recovery of the glitch is due to the elastic motion, since the plastic flow is hard to recover.
The sphere suffering elastic motion would also be fractured, which would make the recovery time scale much larger than the spin-up timescale.
Moreover, we further assume that the stress released by the plastic flow would be restored after glitch, and the unrestored stress is released as the result of the persistent change of oblateness.

Under such starquake scenario, we derive the relation between the recovery coefficient $Q$ and glitch size $\Delta \Omega_g/\Omega$, which is consistent with observational data under reasonable parameters.
We also derive the relation between the time interval of two successive glitches and the released stress, and our results are consistent with the data of the Crab and the Vela pulsars.
Although the real process of starquake is certainly complicated, and the simplified picture cannot explain all data of glitches, it seems qualitatively reasonable and could be the first step in establishing a more sophisticated description about glitches.

Our results could have implications about the oblateness of strangeon stars.
Fig~\ref{fig_Q} indicates that the increase of oblateness $\epsilon^*-\epsilon_m$ reflecting the elastic recovery has the maximum value in the range from $\sim 10^{-7}$ to $\sim 10^{-6}$, which may imply that the actual oblateness of the strangeon stars could be small ($\gtrsim10^{-7}$).
In addition, Fig~\ref{fig_tq} indicates that the persistent decrease of oblateness $|\Delta \epsilon|= \epsilon_g-\epsilon^*$ of the Crab pulsar is larger than that of the Velar pulsar by nearly one order of magnitude, which implies that the actual oblateness of the Crab pulsar might be larger than that of the Vela pulsar by about one order of magnitude.

Here we consider the recovery of glitches to be the result of elastic motion.
However, the recovery process should be complicated and depend on many factors, such as the mass, the magnetic field and spin frequency.
Moreover, the parameter $a$ in Eq.(\ref{Q}) may also different for different pulsars.
Obviously, it is difficult to construct a model to account for all the data in Fig.\ref{fig_Q}, so here we only use the data to find the possible range of the values of $a$ in the simplified model.
We hope that the model as well as the implications about the oblateness of strangeon stars could be improved in the future work.

 Only the bulk-invariable starquakes are demonstrated by which we could understand the behaviors of glitches without significant energy releasing.
Glitches detected from AXPs/SGRs, on the other hand, could come from the bulk-invariable starquakes as demonstrated in~\cite{Zhou2014}.
What kinds of starquakes happen in one pulsar could depends on the spin frequency, mass, and/or the configuration of magnetic field.
For example, young pulsars would undergo the bulk-invariable starquakes as the change of volume could not be significant when the stars are fast rotating, while the bulk-variable starquakes would happen in pulsars with low spin frequencies and/or large masses.

It is worth noting that, we have proposed a possibility that small size glitches could be the result of the accretion of strange nuggets (the relics of cosmological QCD phase transition) by pulsars~\citep{LX2016}.
The small size glitches referred to are defined by the mount of energy released in glitches, including those which have smaller $t_q$ than that predicted in starquake model.
Taking this possibility into account, the data points below the horizontal lines could be the result of such glitch trigger mechanism.

Furthermore, we can infer that, starquakes of strangeon stars would have different glitch sizes and recovery behavior, depending on whether they are dominated by plastic flow or elastic motion.
As we have mentioned before, the two typical glitching pulsars, the Crab and the Vela pulsars, show very different glitch behaviors.
Glitches of the Crab pulsar have small sizes and large recovery coefficient, whereas glitches of Vela pulsar usually have large sizes and small recovery coefficient.
If we take this difference as the a real phenomenon instead of the an observational effect, it could imply that starquakes of the Crab pulsar are dominated by elastic motion, and that of the Vela pulsar are dominated by plastic flow.
Although our simplified model at this stage could not explain the above implication, it seems interesting and worth exploring.

It is also worth mentioning that, we discuss the starquake process based on the scenario described in Baym \& Pines (1971), and assume that the strain of strangeon stars is only due to spinning down.
The decaying magnetic field may also produce the strain, but this effect is not considered in this paper.
Although the role of the magnetic field in producing the strain could be less significant than that of spinning down, it is worth exploring in the future.

Glitches are important for us to understand the interior structure of pulsar-like compact stars.
However, it is a challenge to quantitatively describe glitch behaviors, no matter in neutron star model or quark star model, since the related physical processes are complicated.
In this paper we try to give a rough description of glitch behaviors of solid strangeon stars, including the glitch sizes, the recovery and the time interval between two successive glitches.
Certainly, the strangeon star model should be improved, and a more sophisticated description about glitches can help us to better understand the state of dense matter at supranuclear densities.

{\bf Acknowledgments:}
We would like to thank useful discussions at the astrophysics group in Xinjiang University and the pulsar group in Xinjiang Astronomical Observatory. This work is supported by the West Light Foundation (XBBS-2014-23), the National Key R\&D Program of China (No. 2017YFA0402602), the National Natural Science Foundation of China (Nos. 11203018, 11503008,1673002, U1531243), and the Strategic Priority Research Program of CAS (No.  XDB23010200).

\end{document}